\documentclass[12pt]{article}  
\usepackage{graphicx}
\usepackage{dcolumn}
\usepackage{bm}

\begin{document}

\title{A numerical study of spectral properties of the area operator
in loop quantum gravity}
\author{G\'abor Helesfai, and Gyula Bene \footnote{E-mail: bene@arpad.elte.hu}\\
Institute for Theoretical Physics, E\"otv\"os University,\\
     P\'azm\'any P\'eter s\'et\'any 1/A, H-1117 Budapest, Hungary\\
E-mail: bene@arpad.elte.hu}

\date{\today}

\maketitle

\begin{abstract} 
The lowest 37000 eigenvalues of the area operator in loop quantum gravity is calculated and studied
numerically. We obtain an asymptotical formula for the eigenvalues as a function of their
sequential number.
The multiplicity of the lowest few hundred eigenvalues is also determined and the smoothed spectral density
is calculated. The spectral density is presented for various number of vertices, edges and SU(2)
representations. A scaling form of spectral density is found, being a power law for one vertex,
while following an exponential for several vertices. The latter case is explained on the basis
of the one vertex spectral density. 
\end{abstract}

%\keyword{loop quantum gravity, area operator, spin networks}

\section{Introduction}
Loop quantum gravity \cite{loop, loop2} aims at quantization of the gravitational field
without relying on a background metric, while respecting the 3+1 dimensional nature of spacetime.
This approach is based on Ashtekar's new canonical
variables \cite{Ashtekar} which reduce the constraints to a polynomial form. Note that the
configuration space variables are connections rather than components of the metrical tensor.

Upon quantization, geometrical quantities like area or volume
become operators acting on the Hilbert space. Area operators are especially interesting, since
their spectral properties rendered possible recently testing the theory, namely, near black holes
correspondence with the Bekenstein-Hawking entropy\cite{Bekenstein} in the quasiclassical limit 
has been established\cite{Krasnov, Rovelli}.

It is a technically important feature of the theory that, by using the so called loop representation,
the Hilbert space is spanned by the spin network states. 
Such a state is characterized by a graph, whose edges are labelled by irreducible SU(2)
representations (or, correspondingly, by integer or half-integer numbers) and whose vertices
lie on a given surface (whose area is to be characterized). Vertices are labelled by zero (unit
respresentation of SU(2)) if gauge invariant states are to be considered (like in this paper).
Edges can be grouped according to their position with respect to the given surface.
Hence, an edge can be of type $u$, $d$ or $t$, corresponding to up, down and tangential
orientation. Irreducible representations on edges emanating from the same vertex and 
being of the same type should be multiplied directly and the product should be 
subsequently decomposed to irreducible representations. A state is labelled also by one of
these irreducible representations (for each vertex and each edge type). Thus, to each vertex
there belongs a triple $j_u$, $j_d$ and $j_t$. These are not quite independent, as it is required that
the decomposition of the direct product $j_u\otimes j_d\otimes j_t$ should contain the
unit representation.

\section{Eigenvalues of the area operator}
In the present paper we study the spectral properties of the area operator numerically.
Actually a lot is known about its eigenvalues and eigenstates analytically.\cite{AL}
Each spin network state is an eigenstate of the area operator.
The eigenvalues are explicitly expressed as (with a sum over the vertices)
\begin{eqnarray}  
S=\frac{l^2_P \beta}{2}\sum_{i=1}^V \sqrt{2j_{u,i}(j_{u,i}+1)+2j_{d,i}(j_{d,i}+1)-j_{t,i}(j_{t,i}+1)}  
\end{eqnarray} 
where $l_P$ stands for the Planck length and $\beta$ for the Immirzi parameter whose value
is argued to be $\frac{\ln 3}{2\pi \sqrt{2}}$ \cite{Dreyer}. Note that for this latter value
the Bekenstein-Hawking black hole entropy can be obtained from the theory.

Our aim here to study the spectrum numerically, i.e., to make it numerically transparent what
the formulae yield. As long as just the eigenvalues are of interest, 
it is enough to specify the triple $(j_{u,i},\;j_{d,i} ,\;j_{t,i})$ for each vertex,
the only restriction being that
\begin{eqnarray}  
|j_{u,i}-j_{d,i}|\le j_{t,i}\le j_{u,i}+j_{d,i}
\end{eqnarray}

Hence, $j_{u,i}$ and $j_{d,i}$ run over all integer and half-integer number,
while  $j_{t,i}$ runs from $|j_{u,i}-j_{d,i}|$ to
$j_{u,i}+j_{d,i}$ in unit steps.
In the actual calculation a maximal value $J$ for $j_{u,i}$ and $j_{d,i}$ was applied and checked subsequently
whether the spectrum changed if $J$ was further increased. Moreover, because of the symmetric
role of $j_{u,i}$ and $j_{d,i}$, it was enough to consider the case when $j_{u,i}\le j_{d,i}$.

\begin{figure}[h]
%\begin{flushleft}
\includegraphics{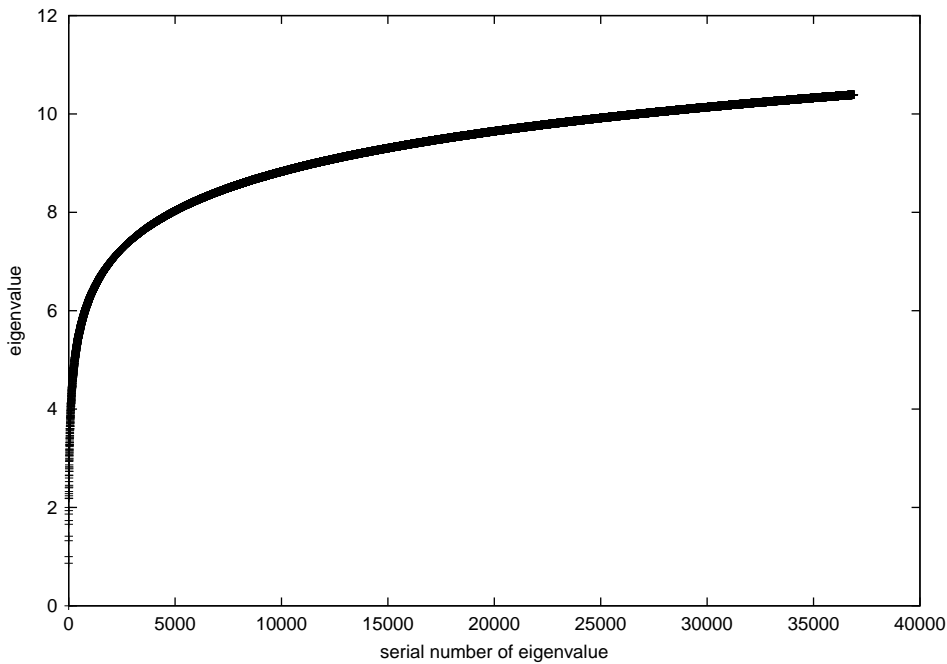}
%\end{flushleft}
\caption{The lowest 37000 eigenvalues versus their sequential number.}
\label{fig1}
\end{figure}

\begin{figure}[h]
%\begin{flushleft}
\includegraphics{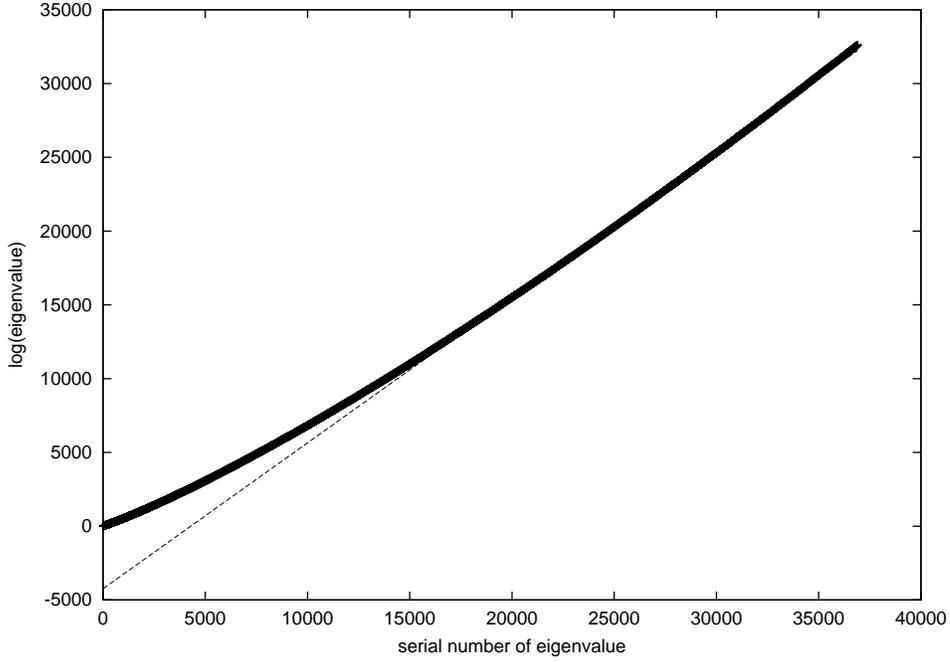}
%\end{flushleft}
\caption{Fitting the lowest 37000 eigenvalues with a logarithmic function.}
\label{fig2}
\end{figure}

After having the eigenvalues, they have been sorted. The eigenvalues versus their sequential number are displayed in Figure \ref{fig1}. 
If plotting the exponential of the eigenvalues, the curve becomes asymptotically a straight line (cf. Fig. \ref{fig2}.). 
Thus, we have asymptotically

\begin{eqnarray}  
S_n=\ln(a\;n+b)\;.  \label{ev}
\end{eqnarray}

with

\begin{eqnarray}  
a=0.931888\pm 0.0002612\\  
b=-2660.66\pm 5.976  
\end{eqnarray} 
 
Certainly, for small sequential numbers $n$ the asymptotical expression becomes meaningless.

The spectrum is discrete, but the mean nearest level spacing decreases
for increasing eigenvalues. It is also clear, that the spectrum is not bounded from above.
The largest nearest level spacing occurs between zero (the smallest eigenvalue) and the next one,
which is $\frac{\sqrt{3}}{2}$\cite{loop}. 

\section{Density of states of the area operator}

In order to get the density of states, one also needs to know the multiplicities of the
eigenvalues. This can be calculated for a given number of vertices, edges and maximal
value $J$. 
If the eigenvalue $S_j$ has the multiplicity $g_j$, then the density of states may be expressed
formally as
 
\begin{eqnarray}  
\rho(S)=\sum_j g_j\delta(S-S_j)
\end{eqnarray}

One may smoothen this density by calculating its convolution with some function $h$
to get

\begin{eqnarray}  
\tilde \rho(S)=\int dS'\;h(S-S')\rho(S')=\sum_j g_j\;h(S-S_j)
\end{eqnarray}

The multiplicities were calculated as follows.
In the one vertex case the irreducible representations on the edges of the same type must be directly
multiplied and subsequently decomposed to irreducible representations again, 
to get the numbers $j_u$, $j_d$ and $j_t$. 
To get the multiplicity associated with these numbers (say, $j_u$), one can
perform the standard procedure of adding angular momenta. 
This is the following. Suppose that the irreducible representations
on the type u edges are labelled by $u_1$, $u_2$, ... $u_N$. I case of one edge,
one has a multiplicity of $2u_1+1$. For two edges one has to decompose
the direct product $u_1\otimes u_2$ into irreducible representations.
The resulting $j$ ranges from $|u_1-u_2|$ to $u_1+u_2$. The corresponding
multiplicity (since a given $j$ appears only once) is again $2j+1$.
For more than two edges one can multiply directly the further irreducible
representations (one by one) with the decomposition obtained in the previous step.
Now, however, at a given level one may get the same irreducible representation $j$
in several way. Therefore, if we are to decompose the product $j_{k-1}\otimes u_k$,
and a resulting irreducible representation is $j_k$ (certainly $|j_{k-1}-u_k|\le j_k
\le j_{k-1}+u_k$), then the multiplicity of $j_k$ has to be increased by
\begin{eqnarray}
\frac{g_{j_{k-1}}}{2j_{k-1}+1}(2u_k+1)
\end{eqnarray}
 
After the multiplicity associated with $j_u$ and $j_d$ has been obtained, the total multiplicity
corresponding to the vertex is 
\begin{eqnarray}
g_{j_u}g_{j_d}(2j_t+1)\;.
\end{eqnarray}
Note that the multiplicity accociated with the type t edges is simply
$2j_t+1$, because $j_t$ has to arise in the decomposition of the product
$j_u\otimes j_d$.

In case of several vertices one proceeds as follows.
Multiplicities corresponding to different vertices must be multiplied together,
while the eigenvalues (i.e., $\sqrt{2j_{u,i}(j_{u,i}+1)+2j_{d,i}(j_{d,i}+1)-j_{t,i}(j_{t,i}+1)}$) 
 add up. If the same eigenvalue arises from different graphs, the corresponding multiplicities add up.
As a consequence, if the density of states is $\rho^1(S)$ for one vertex,
it will be
\begin{eqnarray}
\int dS_1 \rho^1(S-S_1)\rho^1(S_1)
\end{eqnarray}
for two vertices, and
\begin{eqnarray}
\int dS_1 \int dS_2 ..\int dS_{V-1} \rho^1(S-S_1-S_2-..-S_{V-1})\rho^1(S_1)\rho^1(S_2)..\rho^1(S_{V-1})
\label{konv}
\end{eqnarray}
for V vertices.

\begin{figure}[h]
%\begin{flushleft}
\includegraphics{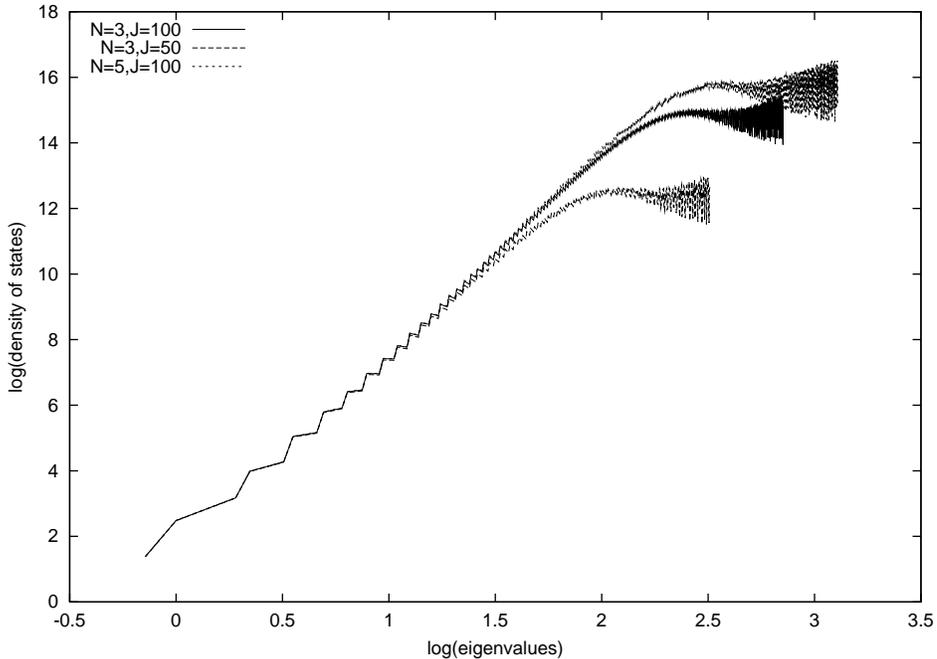}
%\end{flushleft}
\caption{Density of states for one vertex.}\label{fig3}
\end{figure}

Numerically we obtained that the one vertex density of states $\rho^1(S)$ follows asymptotically a power law (cf. Fig. \ref{fig3}),
namely,
\begin{eqnarray}
\rho^1(S)\propto S^{6.2}\;,\label{g1}
\end{eqnarray}
roughly independently of $J$ and $N$ (maximal number of edges).

\begin{figure}[h]
%\begin{flushleft}
\includegraphics{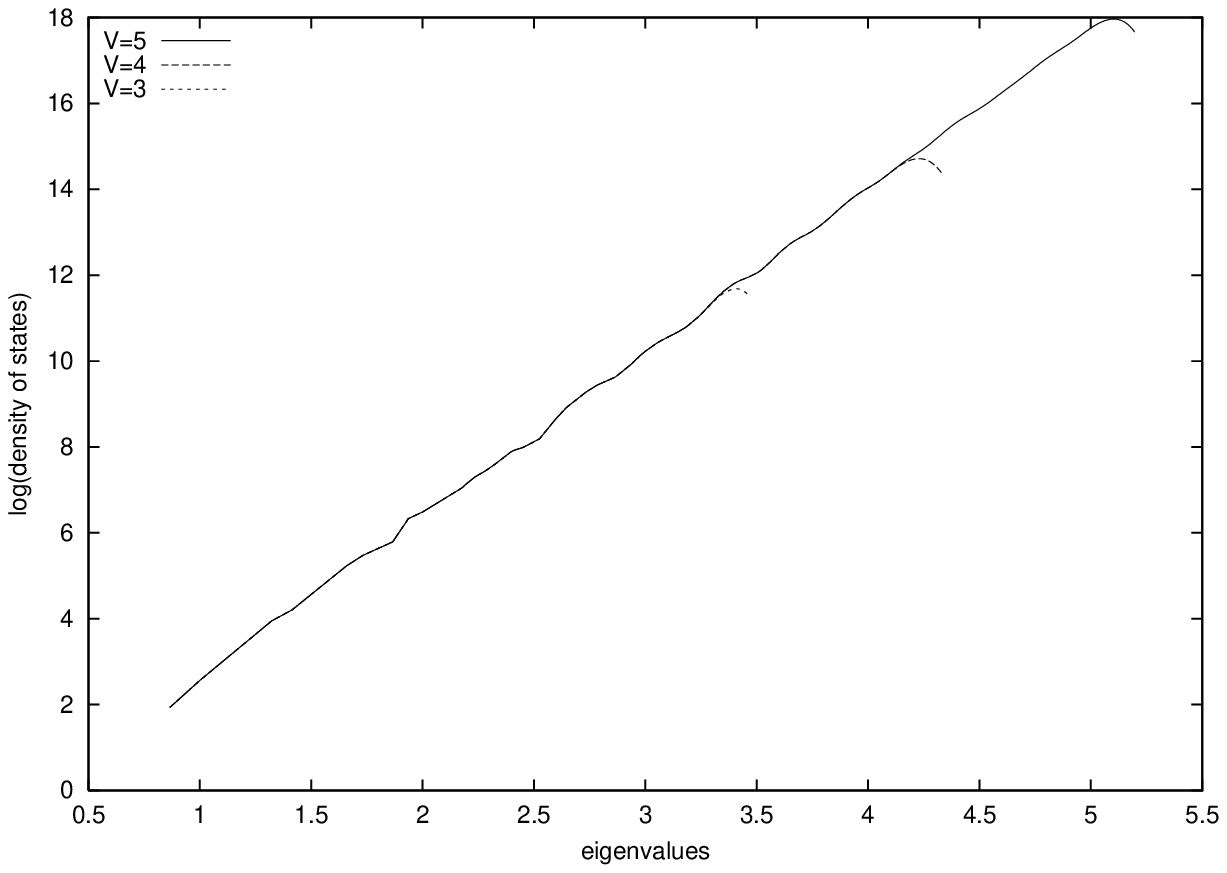}
%\end{flushleft}
\caption{Density of states for several vertices.}\label{fig4}
\end{figure}

For several vertices we obtained that the density of states is asymptotically an exponential (cf. Fig. \ref{fig4}), 
\begin{eqnarray}
\rho^V(S)\propto \exp(3.72 S)\;.\label{gV}
\end{eqnarray}
Eq.(\ref{gV}) looks a bit contradictory in view of Eq.(\ref{konv}), since a convolution of power laws
yields power law again. It turns out, however, that one has to take into account the discreteness
of the spectrum, too. We checked this in terms of a simple model of the one vertex density of states.
This contained an isolated zero eigenvalue, followed by a gap, and then by a continuous power law
behavior. Taking the convolution of this model density of states, we were able to reproduce
the approximately exponential behavior of $\rho^V(S)$.

Note that the proportionality constants did depend on $J$ (maximal index of irreducible representations
on an edge) and $N$ (maximal number of edges emanating from a vertex), but the exponents did not.

\section{Summary}
We presented a numerical study of the spectrum of the area operator. Especially,
we calculated the lowest 37000 eigenvalues, and found numerically the logarithmic asymtotic formula
(\ref{ev}). For the density of states we obtained the power law (\ref{g1}) in case of one vertex,
and the exponential law (\ref{gV}) in case of several vertices. We explained the connection
between these two cases. It is remarkable that the spectral properties are rather
independent of $J$ and $N$ (even of $V$), exhibiting a kind of scaling behavior.
Another interesting observation is that in the general (i.e. $V>1$) case the density of states
is exponential, making possible a formulation of thermodynamics, but leading to an upper bound
for the possible temperatures.
Let us note that no constraints has been used throughout.

\end{document}